# Study of infrared scintillations in gaseous and liquid argon – Part I: methodology and time measurements

A. Bondar,[a,b] A. Buzulutskov,[a,b,*] A. Dolgov,[b] A. Grebenuk,[a] E. Shemyakina[a,b] and A. Sokolov[a,b]

[a] *Budker Institute of Nuclear Physics SB RAS, Lavrentiev avenue 11, 630090 Novosibirsk, Russia*
[b] *Novosibirsk State University, Pirogov street 2, 630090 Novosibirsk, Russia*
*E-mail:* `A.F.Buzulutskov@inp.nsk.su`

ABSTRACT: A methodology to measure Near Infrared (NIR) scintillations in gaseous and liquid Ar, using Geiger-mode APDs (GAPDs) sensitive in the NIR and pulsed X-ray irradiation, is described. This study has been triggered by the development of Cryogenic Avalanche Detectors (CRADs) with optical readout in the NIR using combined THGEM/GAPD multiplier, which may come to be in demand in coherent neutrino-nucleus scattering and dark matter search experiments. A new approach to measure the NIR scintillation yield at cryogenic temperatures has been developed, namely using GAPDs in single photoelectron counting mode with time resolution. The time structure of NIR scintillations and their light yield were measured both for primary scintillations and that of secondary at moderate electric fields (electroluminescence), in gaseous and liquid Ar.

---

[*] Corresponding author.

# Contents

## 1. Introduction

The fact that noble gases can scintillate in the near infrared (NIR) region has been known since the end of 80-ies [1]. This fact was rediscovered at the beginning of the century [2],[3],[4],[5],[6]. Recently there has been a revival of interest in NIR scintillations in noble gases [7],[8],[9], in particular due to the advent of Geiger-mode APDs (GAPDs, [10]) which may have high sensitivity in the NIR [11]. The latter property is employed in so-called Cryogenic Avalanche Detectors (CRADs, [12]) with optical readout in the NIR using combined THGEM/GAPD multiplier, operated in two-phase Ar and Xe. This optical readout technique consists of recording avalanche-induced scintillations emitted from the holes of the thick Gas Electron Multiplier (THGEM, [13]) either in the NIR [14],[15] or VUV [16],[17], using uncoated or Wavelength Shifter (WLS)-coated GAPDs respectively. Such detectors are aimed at reaching single electron sensitivity at low noise, i.e. at having ultimately low detection threshold for primary ionization. Accordingly, they may come to be in great demand in rare-event experiments with low energy deposition such as those of coherent neutrino-nucleus scattering and dark matter search [12].

In gaseous Ar, NIR scintillations were attributed to transitions between the atomic states of the Ar ($3p^5$ 4p) and Ar ($3p^5$ 4s) configurations [1],[2]. In contrast, the emission spectrum of liquid Ar in the NIR is continuous [7],[8]; its emission mechanism has not been yet clarified. Little was known however about the absolute NIR scintillation yield in noble gases: practically nothing about that in Ar and only the lower limit for primary scintillations in Xe ($\geq 21\times10^3$ photon/MeV) [3],[4]. For the first time the NIR scintillation yield in gaseous and liquid Ar was measured in our laboratory both for primary scintillations and those of secondary at moderate electric fields (electroluminescence), at cryogenic temperatures; the first results were published in a short paper [9].

Here we present more elaborated papers on NIR scintillations, studied in gaseous and liquid Ar at cryogenic temperatures. In Part I (the present paper), a methodology to measure NIR scintillations using GAPDs sensitive in the NIR, is described; in particular, the results on the light yield measurement procedure and NIR scintillation time structure are presented. In Part II [18], the experimental results on the light yield of primary and secondary NIR scintillations are presented; possible applications of NIR scintillations in rare-event experiments and ion beam radiotherapy are also discussed.

It should be remarked that any measurement of the absolute scintillation yield is generally a difficult task, since one has to correctly determine the number of photons emitted in a scintillation flash. In the present work we applied a new approach to perform such measurements, namely using GAPDs at cryogenic temperatures in a single photoelectron counting mode with time resolution. Here the cryogenic temperatures provide a superior GAPD performance as compared to that at room temperature [14],[19],[20],[21]: the noise-rate is considerably reduced, while the amplitude resolution and the maximum gain can be substantially increased.

## 2. Experimental setup

Fig. 1 shows the experimental setup; as concerns cryogenics it was similar to that described elsewhere [14]. It consisted of a 9 l cryogenic chamber filled with either gaseous or liquid Ar. Ar was taken from the bottle with a specified purity of 99.998% ($N_2$ content <0.001%); during cooling procedures it was additionally purified from oxygen and water by Oxisorb filter [22], providing the electron life-time in the liquid above 20 μs [23] which corresponds to the oxygen-equivalent impurity content below 20 ppb. The measurements were conducted at cryogenic temperatures: either in gaseous Ar at 87-163 K or in liquid Ar at 87 K.

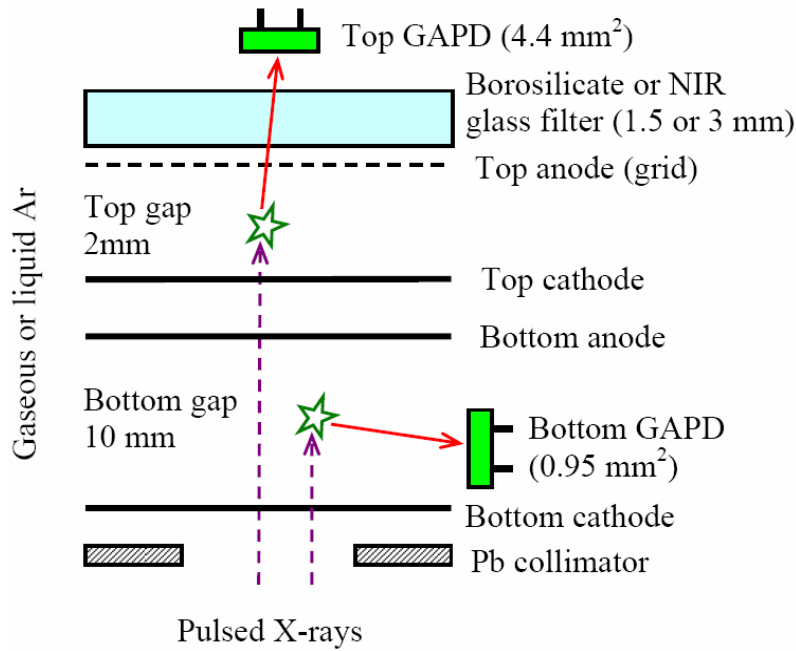

Fig. 1. Schematic view of the experimental setup to study scintillations in the NIR and visible region in gaseous and liquid Ar (not to scale).

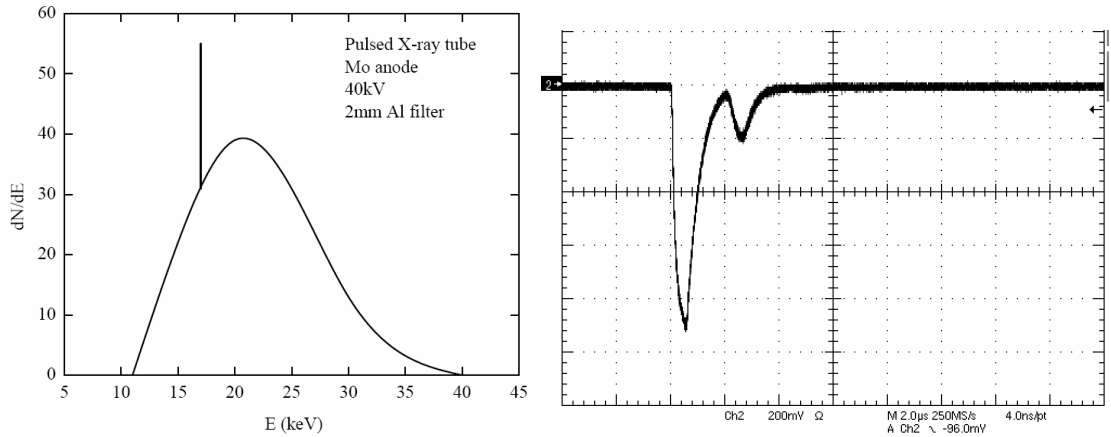

Fig. 2. Characteristic properties of the pulsed X-ray tube with Mo anode used in the present study (0,11BSV7-Mo, [24]): shown are the calculated X-ray spectrum after passing through the Al windows at the bottom of the cryogenic chamber (left) and the X-ray pulse as measured with a BGO scintillation counter (right). The time scale in the right figure is 2 μs/div.

The signals in the cryogenic chamber were induced by X-rays from a pulsed X-ray tube with Mo anode (0,11BSV7-Mo, [24]). The tube was operated at a voltage of 40 kV and anode current of 2.5 mA in a pulsed mode with a frequency of 240 Hz. The latter was provided by a gating grid which was gated by a dedicated pulse generator. The pulse generator also provided a trigger for reading out the data. The chamber was irradiated from outside practically uniformly across the active area, through a lead collimator and aluminium windows at the chamber bottom, defining a cylindrical X-ray conversion region of a 20 mm diameter. The calculated energy of the incident X-rays were within 15-40 keV in gaseous Ar (see Fig. 2, left) and 30-40 keV in liquid Ar. In the latter case, the X-ray absorption in the dead zones at the chamber bottom, filled with liquid Ar, was taken into account.

The X-ray pulse had a sufficient power to provide measurable ionization charges, normally having values of several thousands and tens of thousands electrons in gaseous and liquid Ar respectively. In addition it was sufficiently fast, having a width of 0.5 μs, to provide a reasonable time resolution in time measurements. This is seen from Fig. 2 (right) showing the X-ray pulse signal as recorded with a BGO scintillation counter. One can also see that the major pulse is accompanied with a secondary pulse of smaller amplitude, delayed by 2 μs, arisen due to design features of the pulse generator. The pulse width in Fig. 2 (right), of about 0.8 μs, is larger than that of the X-ray pulse (0.5 μs) due to the contribution of the BGO scintillation decay time (0.3 μs).

To cross-check the correctness of the scintillation yield measurement procedure, several measurement runs were conducted in a variety of ways: using two different GAPDs, at different solid angles, with and without a NIR transmission optical filter. In particular, the scintillation and ionization signals were recorded in two different parallel-plate gaps, top and bottom, with an active area of 30×30 mm$^2$ each and thickness of 2 and 10 mm respectively (see Fig. 1). The gaps were formed by thin-plate electrodes made of copper-clad G10 and a wire grid as shown in Fig. 1.

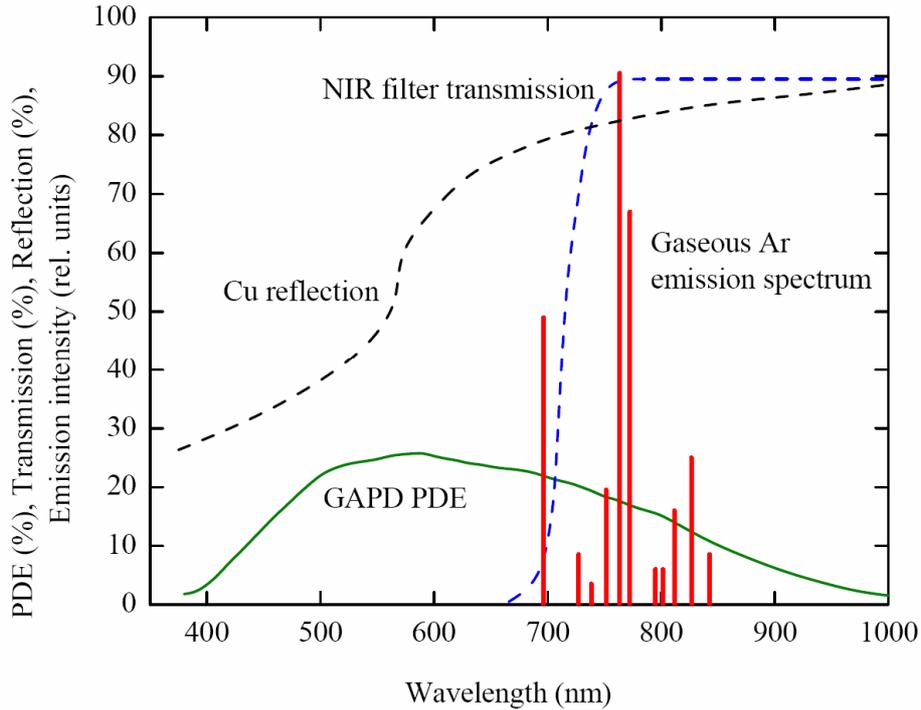

Fig. 3. Spectral characteristics in the visible and NIR region of the gaseous Ar emission due to proton beam-induced scintillations [1], Photon Detection Efficiency (PDE) of the GAPD "CPTA 149-35" [25],[26],[27], transmission of the NIR filter KS-19 [28] and reflection of Cu [29].

The gaps were viewed by respectively "top" and "bottom" GAPDs as shown in Fig. 1 (manufactured by CPTA company [25]), namely by "CPTA 149-35" and "CPTA 143-22" having 4.41 and 0.95 $mm^2$ sensitive area, respectively. The GAPDs were sensitive in the wavelength range of 400-1000 nm: see Fig. 3 showing spectral characteristics of the GAPD and that of gaseous Ar emission. One can see that in gaseous Ar emission region, namely between 690 and 850 nm [1], they had a relatively high Photon Detection Efficiency (PDE): PDE~17% on average [25],[26],[27]. It is essential that the measurements were carried out at cryogenic temperatures where GAPDs have a superior performance as compared to room temperature, in terms of the noise rate, amplitude and single-photoelectron characteristics [20].

The top GAPD was placed at a distance of either 2 or 3.5 mm from the top anode. In front of it either a borosilicate glass filter, transparent to the visible and NIR light, or a NIR glass filter (KS-19, [28]) transparent to the NIR above 700 nm (see Fig. 3), was placed . The bottom GAPD had no optical filter; it took a side view of the gap, at a distance of 25 mm from the gap center (Fig. 1).

The scintillation signal was read out from either the top or bottom GAPD via ~1 m long cable connected to a fast amplifier (CPTA, [25]) with 300 MHz bandwidth and amplification factor of 30. The charge (ionization) signal was read out from the anode electrode of either the top or bottom gap using a charge-sensitive amplifier with a shaping time of 10 μs. Both amplifiers were placed outside the cryogenic chamber. The signals were digitized and memorized for further off-line analysis with a TDS5032B digital oscilloscope, by the trigger provided by the X-ray tube pulse generator.

## 3. Measurement procedure

### 3.1 Scintillation yield

In the present work we applied a new approach to measure the scintillation yield of both primary and secondary scintillations, namely using pulsed X-ray irradiation and GAPDs in a single photoelectron counting mode with time resolution. For each X-ray pulse, the scintillation signal amplitude and that of the charge signal were measured in a given gap, i.e. in the top or bottom gap. The conceptual diagram of the measurement procedure for the scintillation yield is presented in Fig. 4.

The scintillation (light) yield ($Y$) is defined as the ratio of the number of photons emitted over full solid angle ($N_{ph}$) to the number of ionization electrons produced in the gap ($N_e$); in our case both quantities are normalized to X-ray pulse. Note that this definition is valid both for primary and secondary scintillations. In the present work $Y$ was obtained from the GAPD photoelectron yield over full solid angle per X-ray pulse ($Y_{pe}$), divided by the average PDE ($<PDE>$).

$$Y = N_{ph} / N_e = Y_{pe} / <PDE> \ . \tag{1}$$

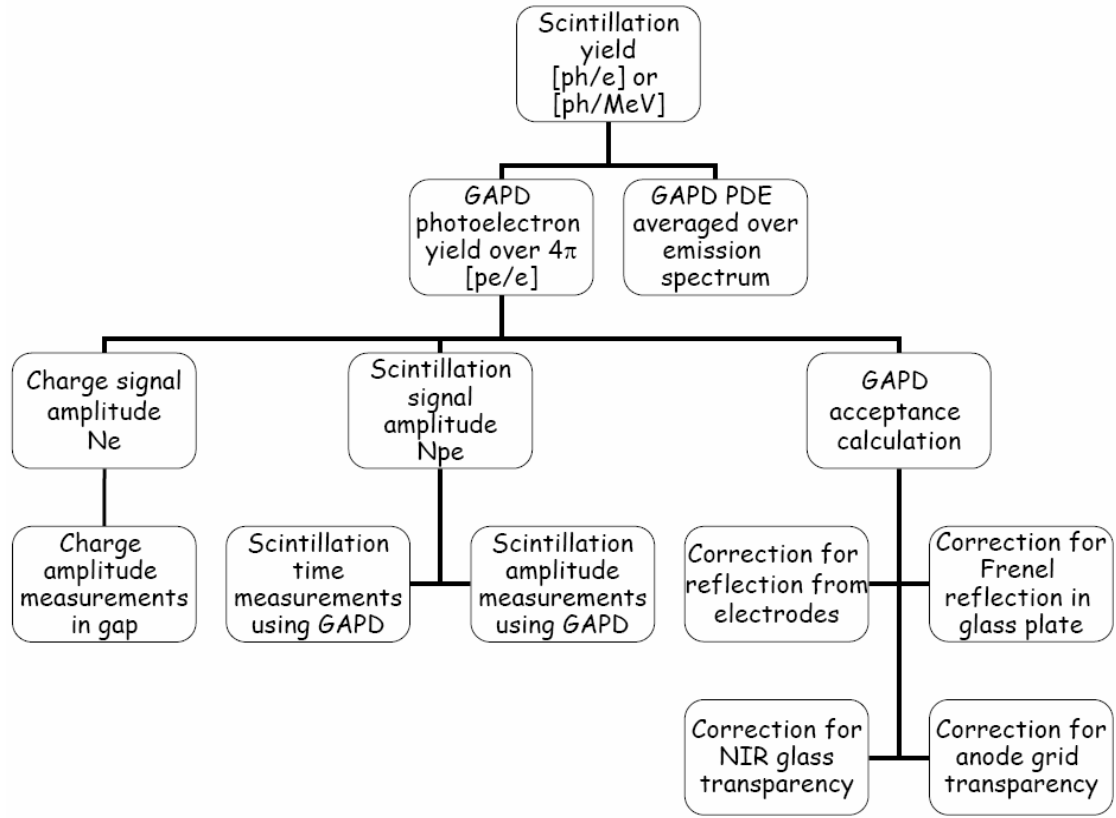

Fig. 4. Conceptual diagram of the measurement procedure for the scintillation yield.

**3.2 Average PDE**

$<PDE>$ was calculated by averaging the PDE spectrum over the Ar emission spectrum passed through the optical filter (see Fig. 3). For that, the GAPD PDE spectrum, provided by the manufacturer [25] and measured elsewhere [26], and the Ar emission spectrum, taken from Refs. [1] and [8] for gaseous and liquid Ar respectively, were used. $<PDE>$ was equal to 17.6% and 16.7% when using the borosilicate glass filter (or no filter) for gaseous and liquid Ar respectively, and to 12.6% when using the NIR filter for gaseous Ar. In the latter case, the Ar emission spectrum was additionally convoluted with the NIR filter transmission spectrum shown in Fig. 3.

It should be noted that the calculation accuracy of $<PDE>$ depends on the validity of the emission spectrum and that of PDE. For the former, one could asses the validity by averaging PDE over the Ar emission spectra obtained by different groups [1],[2],[8], shown Fig. 5. The relative difference in <*PDE*> values thus obtained amounts to 0.1, i.e. it is not that big. Regarding the PDE spectrum, it was validated in special measurements conducted in our Institute using a dedicated Hamamatsu Si photodiode as reference [27]: the NIR part of the PDE spectrum presented by the manufacture [25],[26] was reproduced with reasonable relative error (below 0.1) up to 800 nm, and was specified for longer wavelengths. Thus we may conclude that the relative error of <*PDE*> was within 0.15.

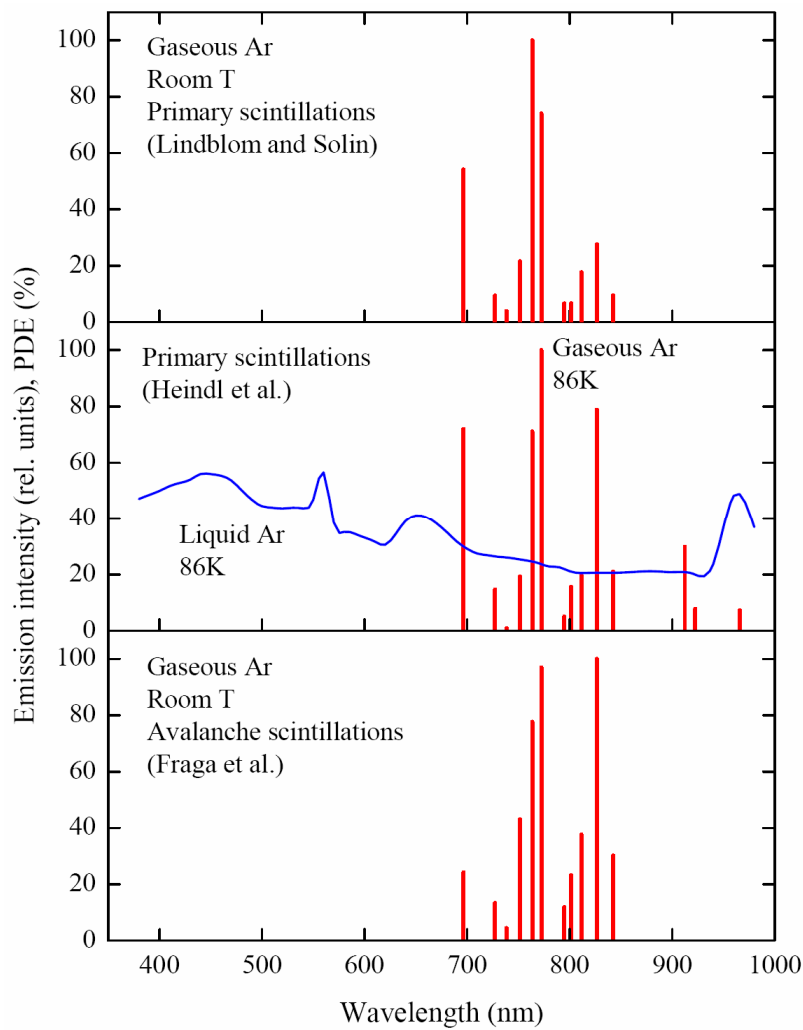

Fig. 5. Emission spectra of gaseous Ar measured by different groups: for primary scintillations - by Lindblom and Solin at room temperature [1] (top) and by Heindl et al. at 86 K [8] (middle); for secondary avalanche scintillations - by Fraga et al. at room temperature [2] (bottom). The emission spectrum of liquid Ar measured by Heindl et al. [8] is also shown (middle).

### 3.3 Photoelectron yield

The next step of the measurement procedure, illustrated in Fig. 4, is the measurement of the GAPD photoelectron yield over full solid angle per X-ray pulse ($Y_{pe}$). It was obtained from the scintillation signal amplitude ($N_{pe}$), divided by the charge signal amplitude ($N_e$), both quantities normalized to X-ray pulse, and the GAPD average acceptance with respect to the X-ray conversion region ($A_{GAPD}$):

$$Y_{pe} = N_{pe} / N_e / A_{GAPD} \quad . \tag{2}$$

Let us consider step by step the procedures to determine these quantities.

$N_e$ is just the number of electrons produced in the gap per X-ray pulse, i.e. the ionization signal amplitude expressed in electrons; it was measured at the anode electrode under an electric field applied across the gap (top or bottom), using calibrated charge-sensitive amplifier.

Similarly, $N_{pe}$ is just the number of photoelectrons per X-ray pulse, i.e. the scintillation signal amplitude as measured by the GAPD (top or bottom) expressed in photoelectrons. As compared to other photodetectors, in addition to their higher sensitivity in the NIR, GAPDs provide a faster single-photoelectron pulse, having well-defined shape (with width of 20 ns) and very small amplitude fluctuation [14],[20]. The latter is illustrated in Fig. 6 presenting GAPD noise signals in liquid Ar at 87 K: one can see a distinct separation between single-, double- and triple-photoelectron signals. These properties result in a remarkable capability of GAPDs to count the number of photoelectrons ($N_{pe}$) contained in a given scintillation signal: by counting single-photoelectron pulses in a time scale corresponding to the scintillation signal duration. To realize this method, the X-ray pulse intensity was reduced down to a level providing a practically non-overlapping of single-photoelectron pulses in the scintillation signal. This was the key element of the measurement procedure.

The number of photoelectrons per X-ray pulse ($N_{pe}$) as recorded by the GAPD, was determined using two different methods: that of scintillation amplitude measurements and that of scintillation time measurements (see Fig. 4).

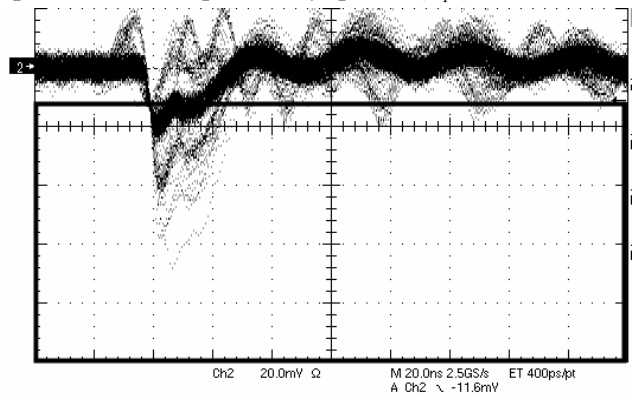

Fig. 6. Noise signals of the top GAPD in liquid Ar at 87 K, at the bias voltage of 44 V. The time scale is 20 ns/div; the vertical scale is 20 mV/div. The heavy line indicates the threshold to hit the time histogram, used in the method of scintillation time measurements.

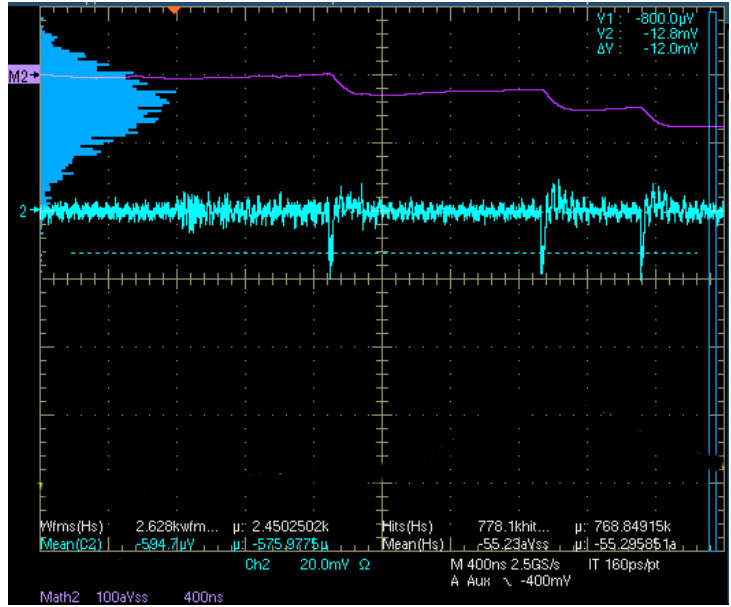

Fig. 7. Method of amplitude measurements to determine $N_{pe}$. Typical X-ray-pulse-induced signal of the top GAPD in liquid Ar at 87 K, at the bias voltage of 44 V, is shown. Here it is composed of 3 successive single-photoelectron (bipolar) pulses (lower trace); the time sweep of their integrated-over-time amplitude is also shown (upper trace). On the left, the distribution of these integrated amplitudes is shown. The time scale is 400 ns/div; the vertical scale for lower trace is 20 mV/div.

### 3.4 The number of photoelectrons determined using amplitude measurements

The first method was similar to that used in ref. [14], where the avalanche scintillation yield was measured in the two-phase Ar CRAD with THGEM/GAPD optical readout. In this method, the original bipolar pulses of the GAPD (after the fast amplifier) were transformed to unipolar ones by integration, corresponding to pulse-filtering with a time constant of ~100 ns. The resulting pulse-area provided the GAPD signal amplitude. This permitted assessing the total amplitudes of long (~10 μs) GAPD signals, consisting of multiple short pulses separated in time, by signal integration over time. This procedure is illustrated in Fig. 7 showing a typical GAPD signal per X-ray pulse, consisting here of 3 successive single-photoelectron (bipolar) pulses; this signal was further filtered and integrated over time, producing the integrated signal amplitude.

To express the integrated amplitude in photoelectrons, the amplitude distribution of GAPD noise signals was used, providing the single-photoelectron (single-pixel) amplitude due to the excellent separation of photoelectron peaks: see Fig. 8. In addition, this distribution also permitted evaluating the cross-talk factor between pixels [14]; this factor was used to correct the final $N_{pe}$ value.

Examples of integrated amplitude distributions of the top GAPD signals, expressed in photoelectrons, are presented in Fig. 9 for gaseous Ar at 159 K and for liquid Ar at 87 K. The distribution mean, after correcting for cross-talks, provides the desired $N_{pe}$ value.

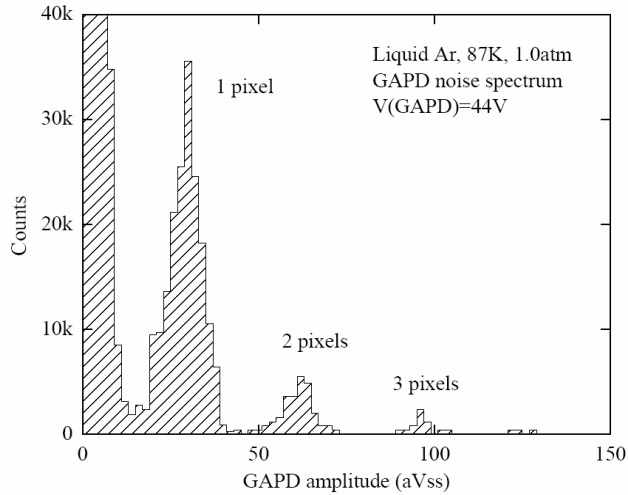

Fig. 8. Typical amplitude distribution of noise signals of the top GAPD in liquid Ar at 87 K, at the bias voltage of 44 V.

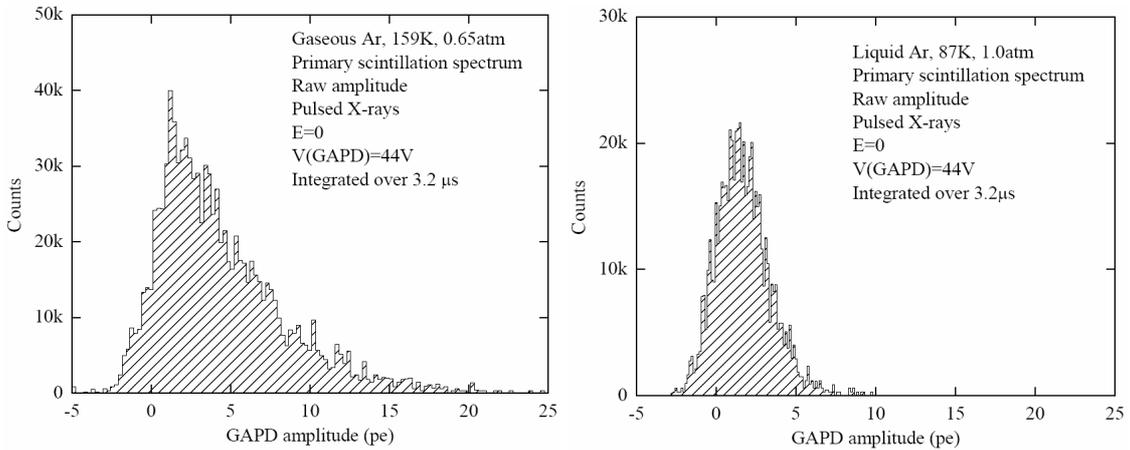

Fig. 9. Typical amplitude distribution of the top GAPD signal per X-ray pulse integrated over 3.2 μs, in gaseous Ar at 159 K and 0.65 atm (left) and in liquid Ar at 87 K (right), at the bias voltage of 44 V. The amplitude is expressed in photoelectrons, not corrected for cross-talks.

### 3.5 The number of photoelectrons determined using time measurements

In the second method, the scintillation time measurements were used; these were also used to determine the time structure of the scintillation signal. Fig. 10 illustrates how this method worked using a TDS-5032B oscilloscope flash ADC: it digitized the pulse and put its part exceeding a given threshold into a time histogram. The threshold was chosen such that the single-photoelectron pulses confidently hit the histogram: see Fig. 6. Thus the time histogram was composed mostly of non-overlapping GAPD single-photoelectron pulses, provided that the GAPD pulses were fast and that the X-ray pulse intensity was appropriately reduced. Accordingly the desired $N_{pe}$ is just the total number of hits in the histogram ($N_{hit}$) divided by the average number of hits per single-photoelectron pulse ($<H>$) and the number of triggers ($N_{trig}$) equal to the number of X-ray pulses:

$$N_{pe} = N_{hit} / <H> / N_{trig} \quad . \tag{3}$$

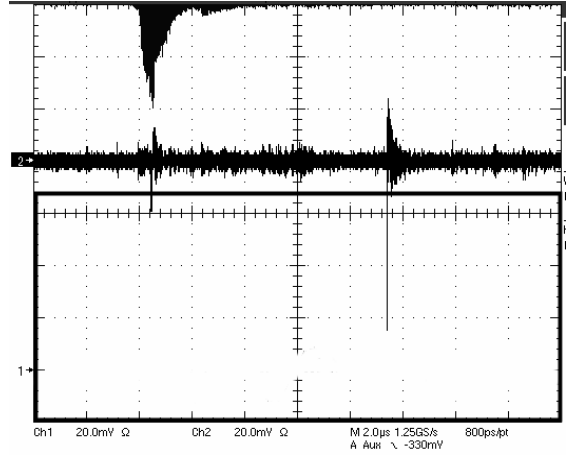

Fig. 10. Method of time measurements to determine $N_{pe}$. Typical X-ray-pulse-induced signal of the top GAPD in gaseous Ar at 159 K and 0.65 atm, at the GAPD bias voltage of 44 V, is shown. Here it is composed of 2 successive (bipolar) pulses, having single- and triple-photoelectron amplitude. On the top, the time histogram is shown; it is composed of hits due to the parts of pulses exceeding the detection threshold indicated by the heavy line. The time scale is 2 μs/div; the vertical scale is 20 mV/div.

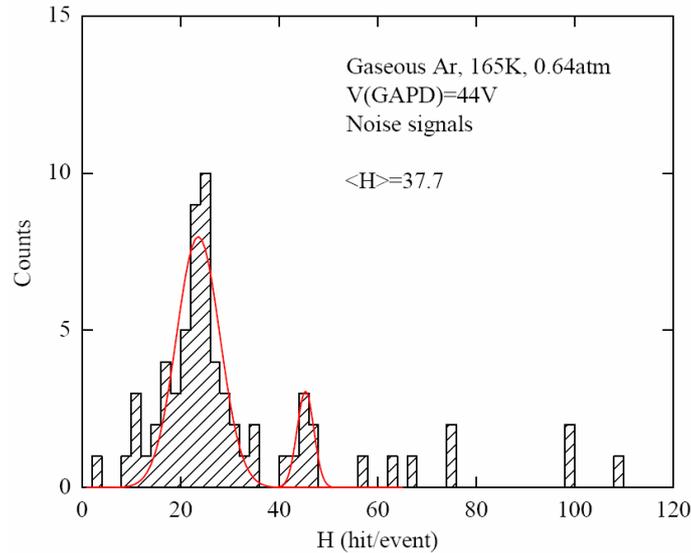

Fig. 11. Typical distribution of the number of hits in the time histogram per GAPD pulse obtained for noise signals of the top GAPD, in gaseous Ar at 165 K, at the GAPD bias voltage of 44 V. The mean of this distribution is equal to $<H>$.

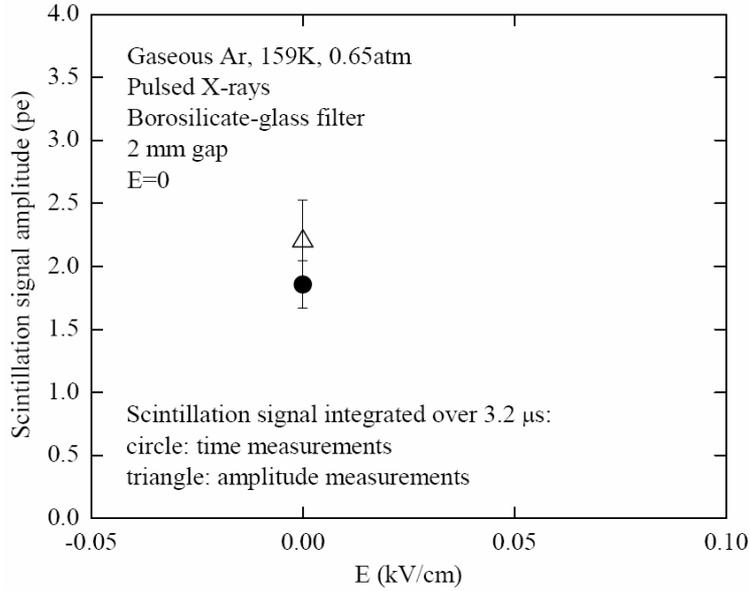

Fig. 12. Comparison of two methods to measure $N_{pe}$. Shown are the scintillation signal amplitudes per X-ray pulse expressed in photoelectrons as measured by the top GAPD in gaseous Ar at 159 K and 0.65 atm, obtained using amplitude measurements and time measurements. The measurement conditions are indicated in the figure.

$<H>$ was determined for each time measurement in special measurement runs using noise signals, conducted at given temperature, GAPD voltage, detection threshold and sampling rate of the flash ADC: see Fig. 11. The figure shows a typical distribution of the number of hits in the time histogram per GAPD pulse. One can see that this distribution is similar to the amplitude distribution of noise signals in Fig. 8, having the single- and double-photoelectron peaks. Consequently, the mean of this distribution, which is by definition is equal to $<H>$, is naturally corrected for GAPD cross-talks.

For the top GAPD when using the borosilicate or NIR glass filter, special measurement runs were conducted in vacuum at a given temperature, i.e. with no signal coming from the gap. This was done to estimate the contribution of the scintillations produced in the glass filter, if any. Though this contribution was found to be negligible, it was also taken into account and subtracted (after appropriate normalization) from the statistics of the major measurement.

In conclusion to this subsection, Fig. 12 compares $N_{pe}$ values obtained by two methods, i.e. by that of amplitude measurements and that of time measurements. One can see that both methods are consistent: they give similar results within experimental errors. This supports the reliability of the methods used. Accordingly in the following, only the method using time measurements will be applied to determine $N_{pe}$ and hence the scintillation yield.

### 3.6 GAPD acceptance

The last step of the measurement procedure was the calculation of the GAPD acceptance $A_{GAPD}$ (see Fig. 4). The latter is actually the average reduced solid angle of the GAPD with respect to the X-ray conversion region, corrected for a number of effects related to light propagation ($<\Delta\Omega_{GAPD}/4\pi>$); it was accurately calculated using Monte-Carlo simulation. In this calculation several important geometrical-optics effects were taken into account.

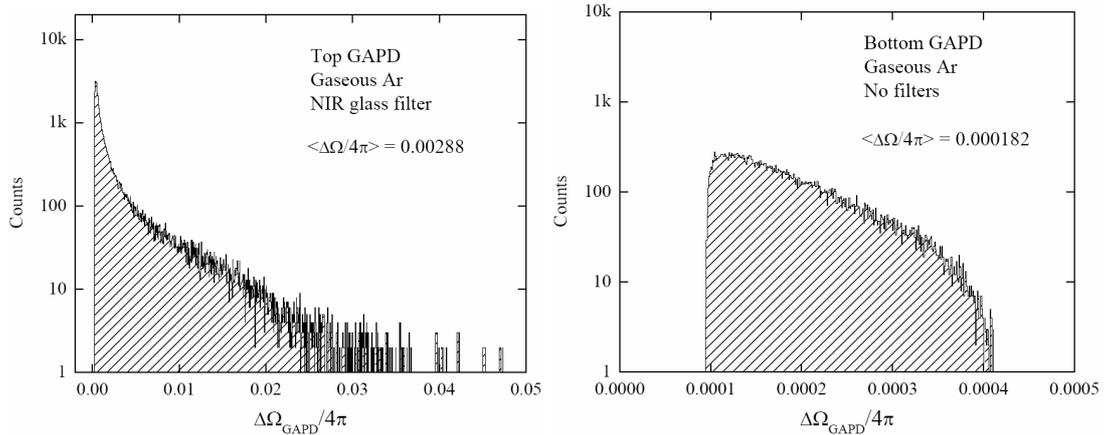

Fig. 13. Distributions of the GAPD reduced solid angle ($\Delta\Omega_{GAPD}/4\pi$) with respect to the X-ray conversion region, in gaseous Ar: for the top GAPD with NIR glass filter (left) and for the bottom GAPD with no filter (right).

In particular for both gaps, the mirror light reflection from the Cu electrodes was taken into account using Cu spectral reflection data in the NIR [29] presented in Fig. 2. In addition for the top gap, the light refraction, Frenel reflections and light absorption (if any) in the optical filter were taken into account; note that all these effects have strong angle dependence. The anode grid transparency of the top gap has angle dependence as well: it was also taken into account.

The results of these calculations are presented in Fig. 13: in particular, the calculated $A_{GAPD}$ values amounted to 0.0029 for the top GAPD (with NIR filter) and to 0.00018 for the bottom GAPD (with no filter), for gaseous Ar. For liquid Ar the values differ slightly due to the effect of liquid Ar refraction. Notice a considerable difference, of a more than an order of magnitude, in acceptances for the top and bottom GAPDs. We will see in the subsequent paper (Part II, [18]) that despite of this considerable difference, the scintillation yields measured with the top and bottom GAPDs will be compatible within measurement errors, strongly supporting the validity of the acceptance calculations.

In conclusion to section 3, regarding the measurement errors, the major sources of uncertainties were classified as follows: the $N_e$ and $N_{pe}$ measurements gave 10-30% and 10-20% respectively (depending on a particular measurement run), the $<PDE>$ calculation gave 15%. In total these resulted in about 20-40% uncertainty in the scintillation yield for a certain measurement run.

## 4. Scintillation time structure and scintillation yield

### 4.1 Fast and slow components

The method of time measurements permitted to determine the time structure of the scintillation signal: this structure is definitely reflected in time histograms considered in section 3.5. In particular, if to look at the time histogram of Fig. 10 (it is vertically inverted), one can see two peaks reflecting the double-peak structure of the X-ray pulse discussed in section 2 (see Fig. 2). The width at half maximum of each peak corresponds to about 0.5 μs, i.e. to the expected width of the X-ray pulse. That means that the scintillation signal has a fast component with a time constant below 0.1 μs; this time scale was expected from the atomic emission scintillation mechanism in gaseous Ar [1],[2]. Such a fast scintillation component prevailed in both gaseous and liquid Ar. Note that the emission spectrum of liquid Ar is continuous, in contrast to gaseous Ar (see Fig. 5), meaning that its emission mechanism is not atomic, i.e. in general it is not obliged to be fast.

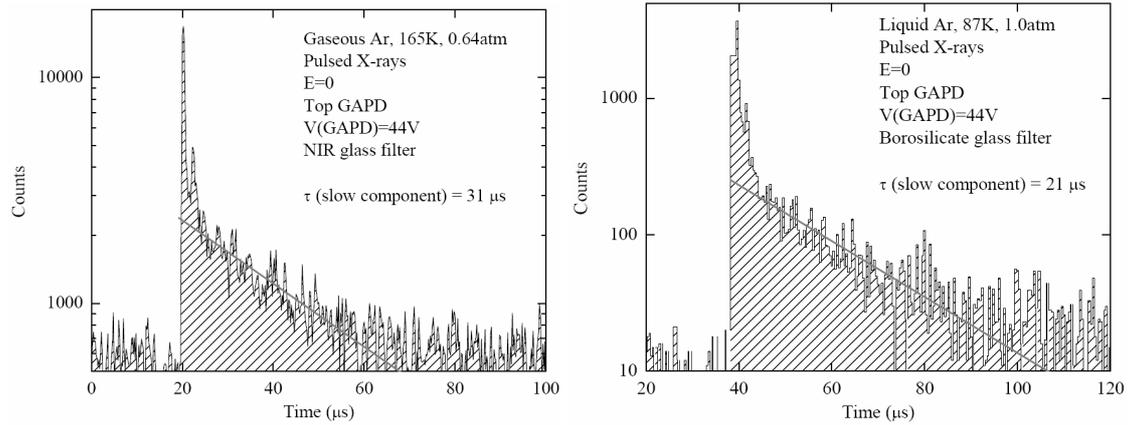

Fig. 14. Time histograms of scintillation signals as measured by the top GAPD in gaseous Ar at 165 K and 0.64 atm with NIR glass filter (left) and in liquid Ar at 87 K with borosilicate glass filter (right). The fast component (the double peak reflecting the time structure of the X-ray pulse) and the slow component (the exponential tail) are seen.

On the other hand in addition to the fast component, the slow scintillation component was observed in both gaseous and liquid Ar, namely the signal tail decaying exponentially with a time constants of about 20-30 μs: see Fig. 14. Its integral contribution to the light yield, in ~50 μs time interval, was comparable to that of the fast component. The origin of the slow component has not been yet understood. Most probably it is caused by slow scintillations of Ar or those of impurities of yet unclear nature; other hypotheses such as delayed X-ray fluorescence, optical filter fluorescence and GAPD after-pulses were tested and practically excluded in supplementary measurements. Accordingly in the following, in the subsequent paper (Part II, [18]), only the fast component contribution to the scintillation yield will be taken into consideration: the slow component contribution was accurately measured and subtracted, using extrapolation procedure similar to that shown in Fig. 14.

### 4.2 Scintillation yield results

The typical result of applying the methods described above to measure the scintillation yield is illustrated in Fig. 15. It shows the scintillation yield in the NIR for the fast component as a function of the electric field in gaseous Ar, at different cryogenic temperatures and pressures. These measurements were carried out in the top gap. The comprehensive discussion of such results will be presented in our subsequent paper [18]. We only indicate here upon the characteristic behavior of the scintillation yield: at all temperatures, the primary scintillations at low electric fields (≤1 kV/cm) are taken over by secondary scintillations at higher fields, i.e. by electroluminescence (or proportional scintillations). The electroluminescence might thus considerably increase the overall scintillation yield in the NIR.

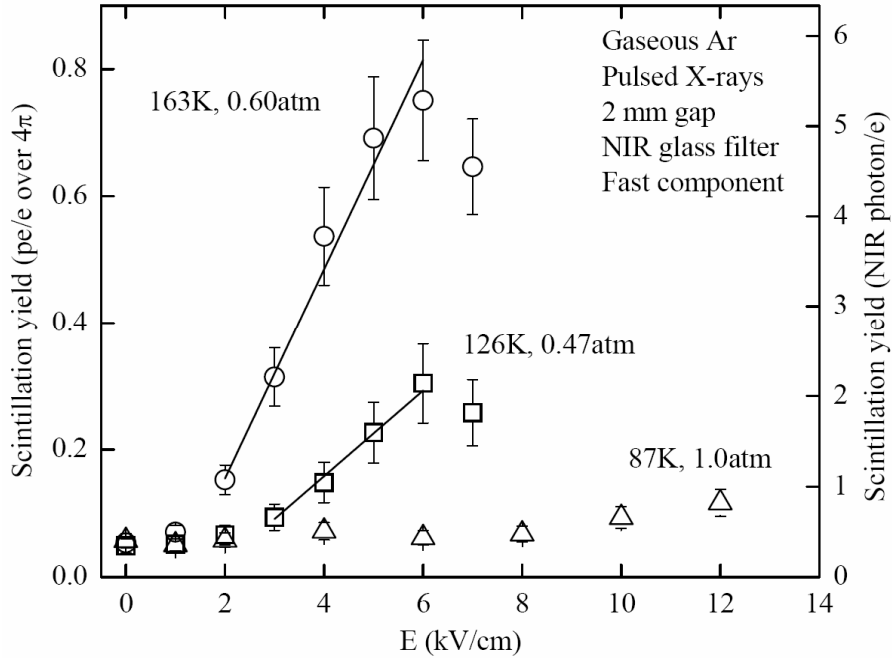

Fig. 15. Scintillation yield in the NIR for the fast emission component as a function of the electric field in gaseous Ar, at different cryogenic temperatures and pressures: at 163 K and 0.60 atm, at 126 K and 0.47 atm, at 87 K and 1.0 atm. The scintillation yield is given in the number of GAPD photoelectrons per ionization electron (left scale) and in the number of NIR photons per ionization electron (right scale), both quantities normalized to full solid angle ($4\pi$). The data were obtained under X-ray irradiation in a 2 mm thick gap (top gap).

## 5. Conclusions

A novel methodology to measure NIR scintillations in gaseous and liquid Ar at cryogenic temperatures has been developed, namely using pulsed X-ray irradiation and GAPDs in single photoelectron counting mode with time resolution. Here the cryogenic temperatures provide a superior GAPD performance as compared to that at room temperature.

The key element of the measurement procedure was based on the remarkable capability of GAPDs to count the number of photoelectrons contained in a given scintillation signal. To perform such counting, two methods were applied: that of scintillation amplitude measurements and that of scintillation time measurements.

The time measurement method permitted to study the time structure of NIR scintillations: fast and slow scintillation components were observed, with time constants below 0.1 μs and of about 20-30 μs respectively, both in gaseous and liquid Ar. The scintillation yield was measured for the fast component, in gaseous and liquid Ar: both for primary scintillations and that of secondary at moderate electric fields (electroluminescence). The results on the scintillation yield, reported recently in the letter [9], will be presented in a more elaborated paper (Part II, [18]).

This study has been triggered by the development of Cryogenic Avalanche Detectors (CRADs) with optical readout in the NIR using combined THGEM/GAPD multiplier (see review [12]). Further studies in this field are in progress in our laboratory.

## 6. Acknowledgements

We are grateful to A. Chegodaev and R. Snopkov for technical support, M. Barnyakov and E. Kravchenko for collaboration and D. Akimov and Y. Tikhonov for discussions. This work was supported in part by the Ministry of Education and Science of Russian Federation, special Federal Program "Scientific and scientific-pedagogical personnel of innovative Russia" in 2009-2013 and Grant of the Government of Russian Federation (11.G34.31.0047).